\documentclass{ws-procs10x7}

\begin{document}

\title{Dileptons and Open Charm: Probes of Chiral 
Restoration}

\author{K. Gallmeister$^1$, B. K\"ampfer$^2$ and S. Zschocke$^{2,3}$}

\address{$^1$ Universit\"at Giessen, D-35392 Giessen, Germany\\
$^2$ Forschungszentrum Rossendorf/Dresden, PF 510119, D-01314 Dresden, Germany\\
$^3$ Technische Universit\"at Dresden, D-01068 Dresden, Germany}

\twocolumn[\maketitle\abstract{
We summarize the status of electromagnetic probes of strongly interacting
matter produced in relativistic heavy-ion collisions at CERN-SPS with
respect to indications of chiral symmetry restoration. Explorative
results for studying the open charm dynamics at BNL-RHIC are presented.  
}]

\section{Introduction}

The dilepton emission rate of thermalized matter reads\cite{Rapp_Wambach,Gale_Haglin}
\begin{equation}
\frac{d R}{d^4 Q} = \frac{\alpha^2}{3 \pi^3 Q^4}{\rm Im} \Pi^R_{\mu \nu} L^{\mu \nu} 
\, f_B(Q \cdot u / T),
\end{equation}
where $Q$ denotes the 4-vector of the pair, $L^{\mu \nu} = Q^\mu Q^\nu - Q^2 g^{\mu \nu}$,
and $f_B$ stands for the thermal (Bose) distribution; $u$ is the 4-velocity of the medium.
The central quantity, $\Pi^R_{\mu \nu}$, is the
retarded photon self-energy in the medium which is related to the
current-current correlator via 
\begin{equation}
\Pi^R_{\mu \nu} = i \int d^4x {\rm e}^{i Qx} \Theta(x_0) 
\langle \langle [J_\mu (x), J_\nu (0)] \rangle \rangle, 
\end{equation}
where $\langle \langle \cdots \rangle \rangle$ means the thermal
average with respect to the medium with temperature $T$ and baryon density $n$;
$J_\mu$ is the electromagnetic current operator.
Also the real photon emission rate is related to 
${\rm Im}\Pi^R_{\mu \nu}$\cite{Gale_Haglin,Peitzmann_Thoma}.
Therefore, the electromagnetic probes are sensitive to the medium properties.
Since the absorption probability of such probes, once emitted, is small
they carry information on the full evolution, in particular also on the early
hot, dense stages in heavy-ion collisions.

\section{Electromagnetic signals at CERN-SPS}
 
The dilepton emission rate, in dilute gas approximation, contains the model-independent 
leading terms
\begin{equation}
\frac{d R}{d^4 Q} =  \frac{4 \alpha^2 f_B}{(2 \pi)^2}
\left\{ \rho^{e.m.} - \epsilon \left( \rho^V - \rho^A \right) 
+ n {\cal T} \right\},
\label{rate}
\end{equation}
where ${\cal T}$ is related to the nucleon matrix element 
of the current-current commutator, 
$\epsilon = T^2/(6f_\pi^2)$ in the chiral limit, and
\begin{eqnarray}
\rho^{e.m.} &=& \frac{1}{12 \pi^2} 
\sigma_{e^+ e^- \to {\rm hadrons}} /
\sigma_{e^+ e^- \to \mu^+ \mu^-}, \nonumber\\
\rho^V &=& \frac{1}{12 \pi^2} 
[\sum_i \sigma_{e^+ e^- \to 2 i \pi}] /
\sigma_{e^+ e^- \to \mu^+ \mu^-}, \nonumber \\
\rho^A &=& \frac{8 \pi m_\tau^3}{G_F^2 \cos^2 \Theta_C (m_\tau^2 + 2 Q^2)(m_\tau^2 - Q^2)}
\nonumber \\
&&\sum_i d \Gamma_{\tau \to \nu_\tau (2i+1)\pi} / d Q^2 
\end{eqnarray}
(cf.\ Ref.\cite{Huang} for details).
\begin{figure}[t]
~\center
\epsfig{file=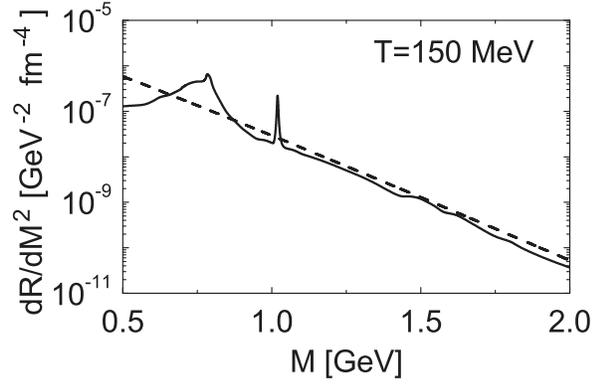,width=215pt,angle=-0}
\caption{Dilepton emission rates of strongly interacting matter
at temperature $T = 150$ MeV. The solid curve is based on Eq.~(\ref{rate}), while
the dashed line is the $\bar q q$ annihilation rate.}
\label{fig1}
\end{figure}
This rate is displayed in Fig.~\ref{fig1} for $n = 0$. It coincides remarkably
well with the Born $\bar q q$ annihilation rate for invariant masses $M > 1$ GeV
in a wide range of $T$; the Born rate in turn is in good agreement with the lattice QCD
evaluation\cite{Wizorke}. A radical point of view is to use the $\bar q q $ rate
as convenient parametrization
in the full range of invariant masses and temperatures
with the arguing that chiral symmetry arguments\cite{Rapp_Wambach}
support a reshuffling of strength in the $\rho - \omega$ region such that also there
a featureless continuum describes the emissivity of strongly interacting matter.  
Moreover, in Ref.\cite{Gallmeister} it has been shown that instead of the use of
the detailed space-time evolution the replacement 
$\int d^4 \, x dR(T, n)/d^4Q \to {\cal N} dR(\langle T \rangle)/d^4Q$ delivers a consistent
description of the dilepton experiments CERES, HELIOS-3, NA38, NA50 at CERN-SPS
with suitably adjusted normalization ${\cal N}$ and space-time averaged temperature
$\langle T \rangle$. 
In Figs.~\ref{fig2} and \ref{fig4}
we show two examples for very recent data in the low-mass region.
In the intermediate-mass region, charm contributions (see below)
become important and the high-mass region is dominated by the Drell-Yan yield.
In particular the forthcoming NA60 data are expected to shed further light
on the role of the charm contribution.  
\begin{figure*}[t]
\epsfig{file=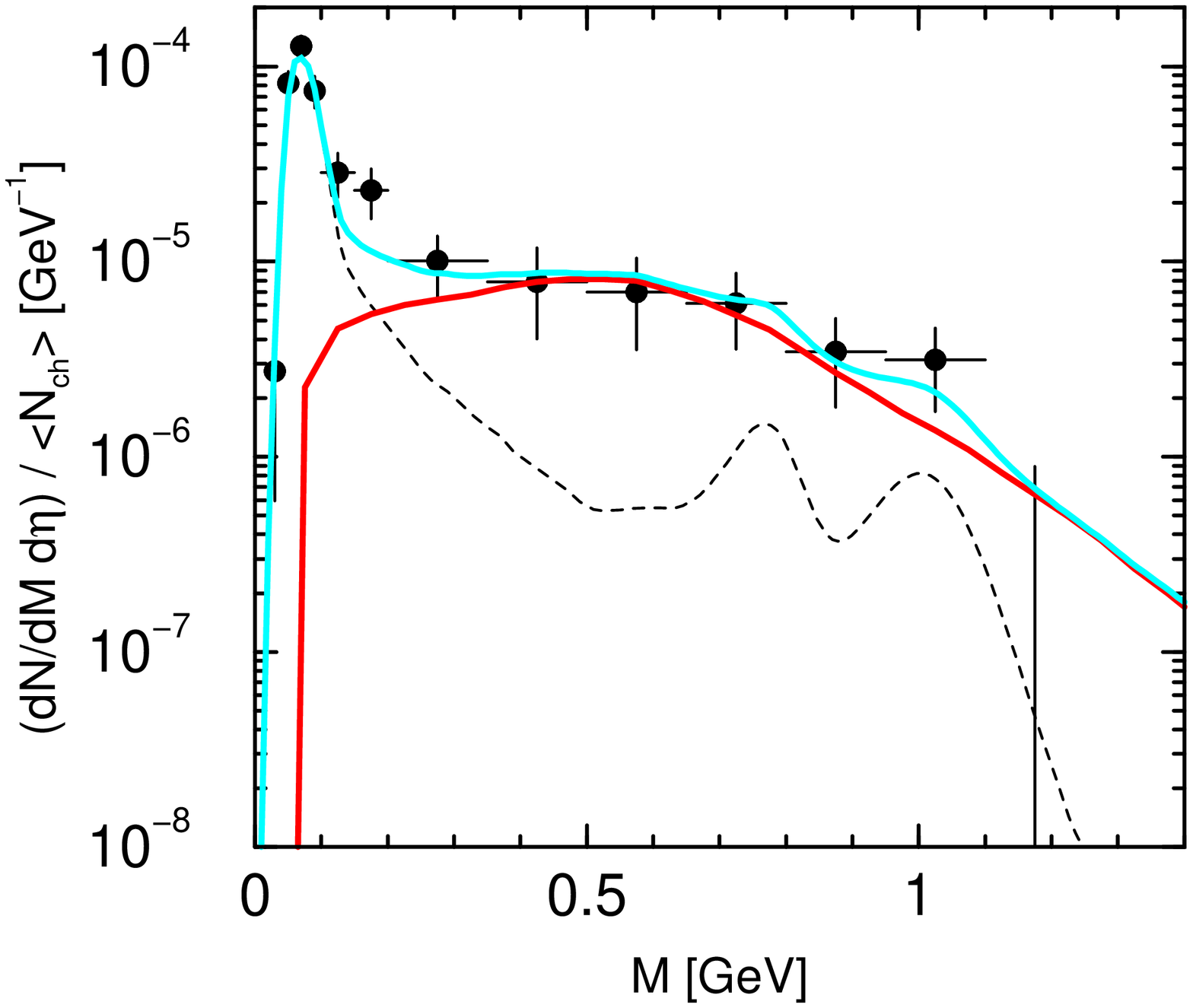,width=150pt,angle=-90}
\hfill
\epsfig{file=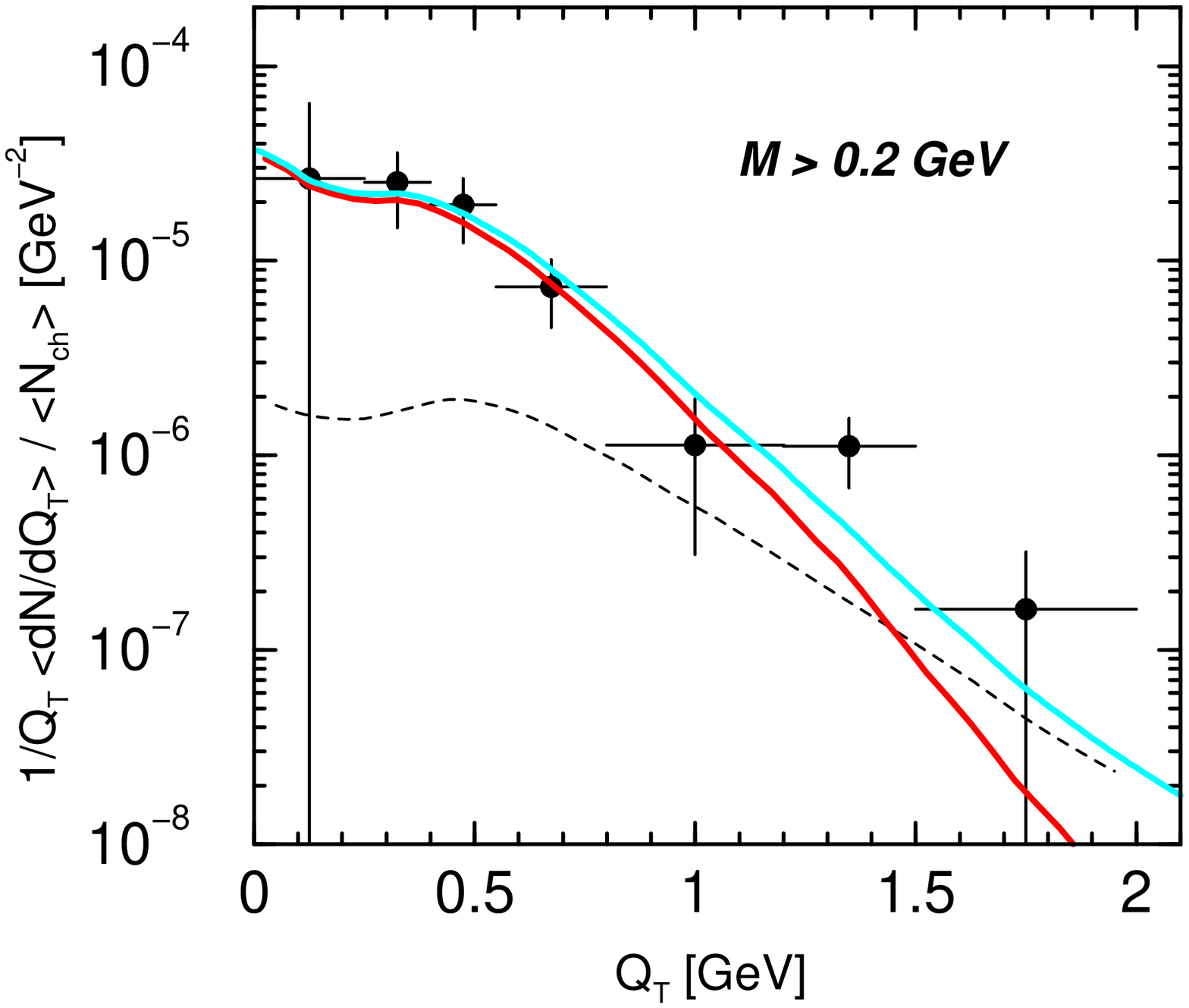,width=150pt,angle=-90}
\caption{Dielectron spectra for the reaction Pb(40 AGeV) + Au
(left panel: invariant mass spectrum, right panel: transverse momentum spectrum,
dashed curves: hadronic cocktail, 
red curves: thermal yield, 
cyan (upper) curves: sum of these contributions). 
Data from Ref.\protect\cite{CERES40}; $\langle T \rangle = 145$ MeV.}
\label{fig2}
\end{figure*}
\begin{figure*}
\epsfig{file=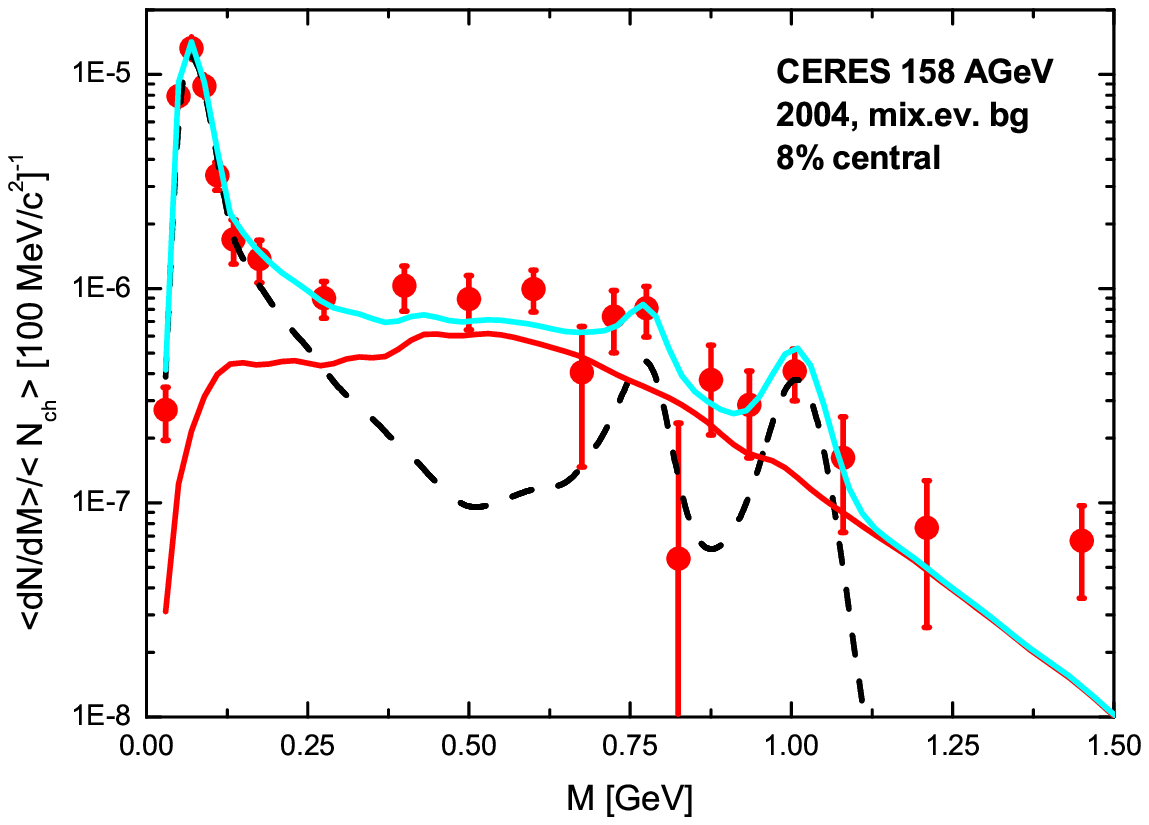,width=180pt,angle=-0}
\hfill
\epsfig{file=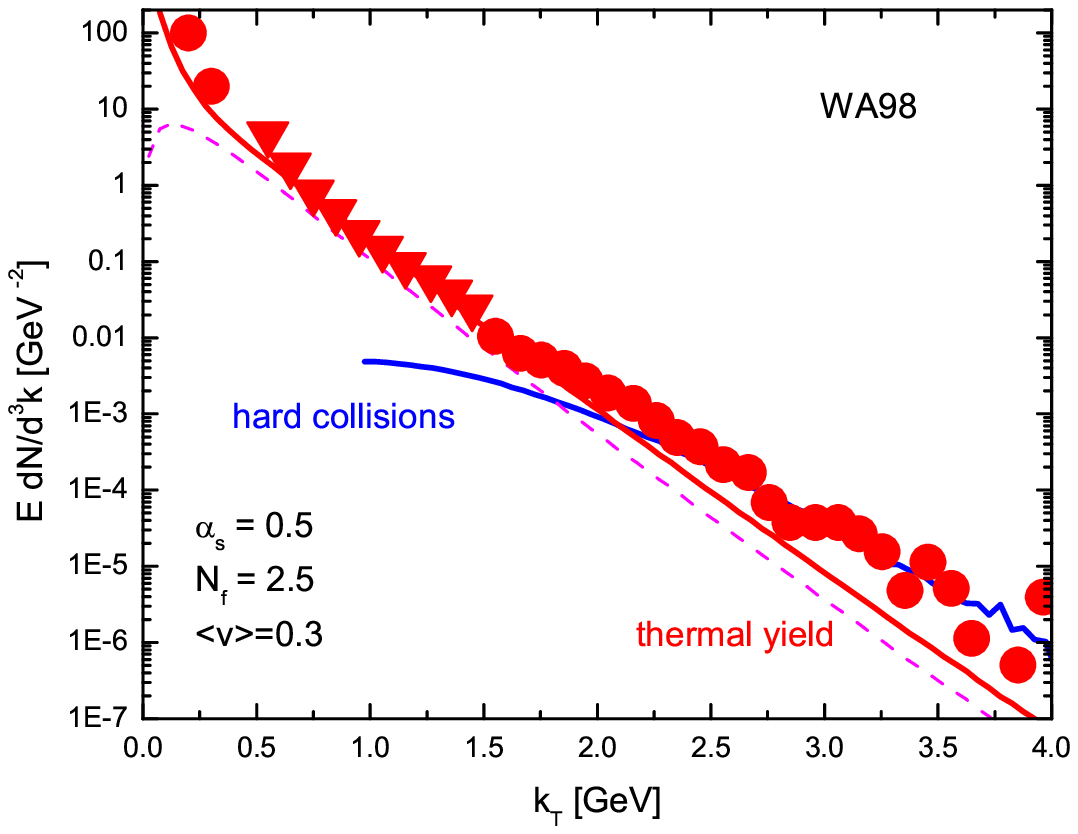,width=170pt,angle=-0}
\caption{Left panel: Dielectron spectrum for the reaction Pb(158 AGeV) + Au
(line codes as in Fig.~\protect\ref{fig2}). Data
from Ref.\protect\cite{CERES158}; $\langle T \rangle = 170$ MeV.
Right panel:
Photon spectrum for the reaction Pb(158 AGeV) + Pb
(blue curve: perturbative hard QCD contribution, dashed curve: lowest-order
thermal yield, red solid curve: thermal yield corrected by resumed 
higher-order\protect\cite{Arnold}).
Data from Ref.\protect\cite{WA98} (error bars are omitted, triangles depict upper limits); 
$\langle T \rangle = 170$ MeV;
averaged transverse expansion velocity $\langle v \rangle = 0.3 c$.}
\label{fig4}
\end{figure*}

When supplementing, in the same sprit,
the Born rate for photons by the collinear enhancement factor\cite{Arnold} 
one gets equally well the description of the recent WA98 data\cite{WA98}, 
see Fig.~\ref{fig4}.

Advanced models (cf.\ Refs.\cite{Rapp_Wambach,Gale_Haglin,Peitzmann_Thoma})
resolve the details of the space-time evolution and make contact to the
equation of state\cite{Peshier} but do not improve the
agreement with data. A particularly important result is that the needed maximum
(i.e., initial) temperature is above the deconfinement temperature.

\section{Charm at RHIC}

\begin{figure}[t]
\epsfig{file=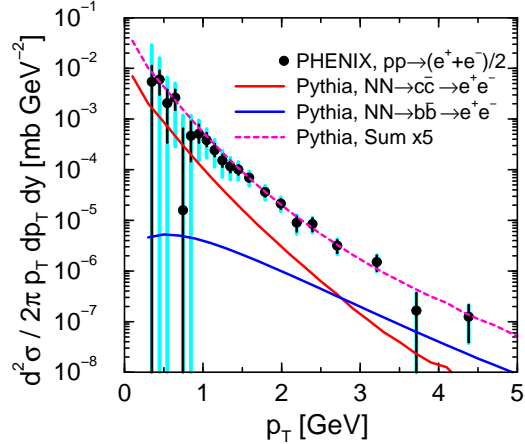,width=175pt,angle=-90}
\caption{Single electron spectrum (data [non-photonic sources]
from Ref.\protect\cite{PHENIX1})
for the reaction p +p at $\sqrt{s} = 200$ GeV compared with results of
PYTHIA (PDF = CTEQ 5L) with default parameters unless hybrid fragmentation and 
$\langle k_\perp^2 \rangle = 2.5$ GeV$^2$, which describe charged hadron spectra
fairly well\protect\cite{Gallmeister_Cassing}. The transverse momentum
spectra\protect\cite{STAR} of $D^0$ and $\bar D^0$ mesons in the reaction
d + Au at $\sqrt{s_{NN}} = 200$ GeV, however, are underestimated 
in normalization and slope for $p_\perp > 6$ 
GeV/c\protect\cite{charm_pT} for these settings.}
\label{fig5}
\end{figure}

To describe the intermediate-mass di\-lepton spectrum at CERN-SPS one must
take into account the contribution of correlated semi-leptonic decays
of open charm mesons\cite{Gallmeister}. At RHIC energies this contribution
is expected to dominate by far. Therefore, the correct treatment of charm is 
important. Experimental data\cite{PHENIX1} on the inclusive single electron
spectra in $pp$ collisions may serve for adjustment purposes. In Fig.~\ref{fig5}
the corresponding spectra are displayed. The charm and bottom cross sections
are 650 and 4.3 $\mu$b, respectively. 
Here we have used a common $K$ factor both for open charm and bottom
of the constant value of 5.

\begin{figure*}
\epsfig{file=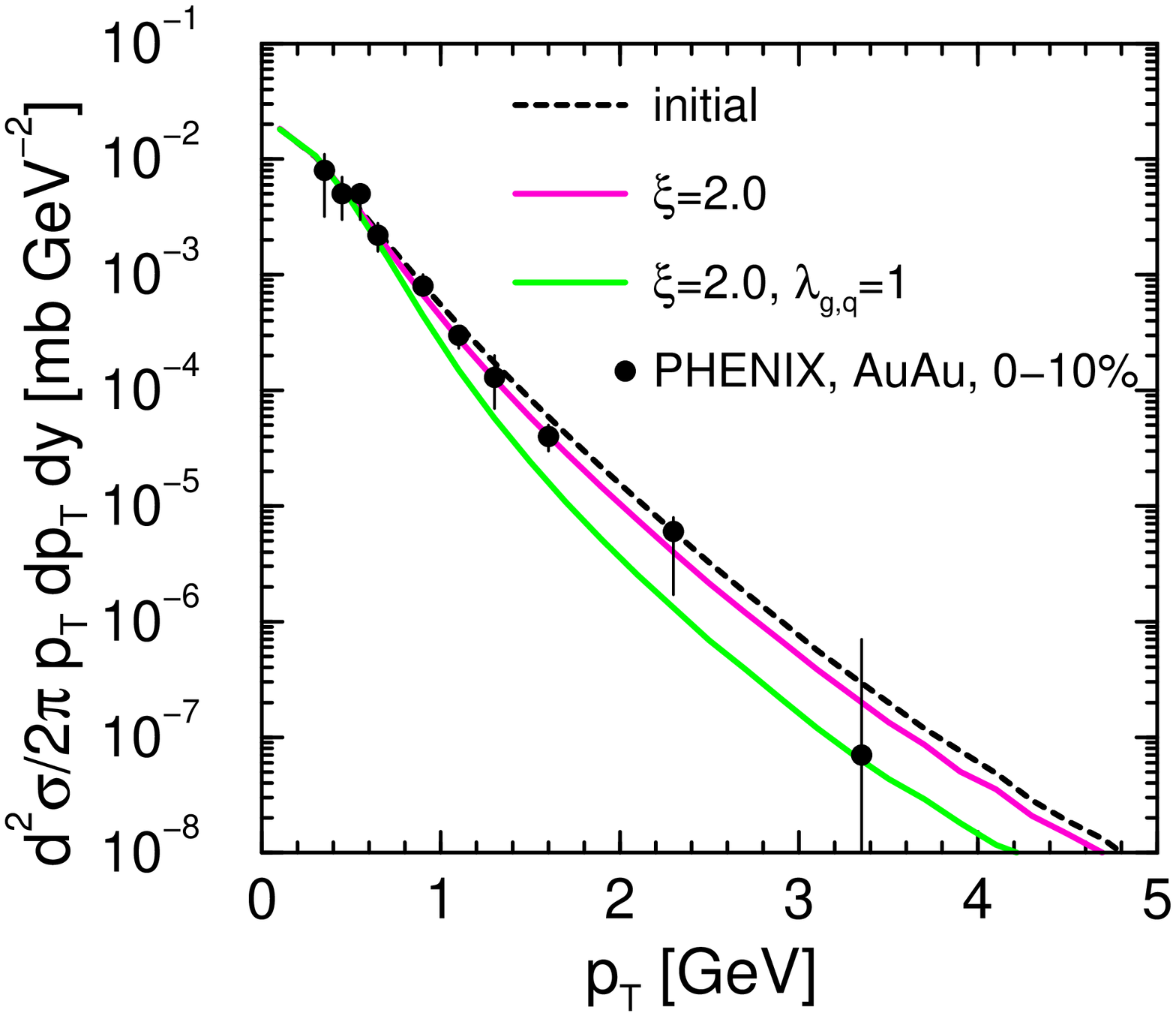,width=160pt,angle=-90}
\hfill
\epsfig{file=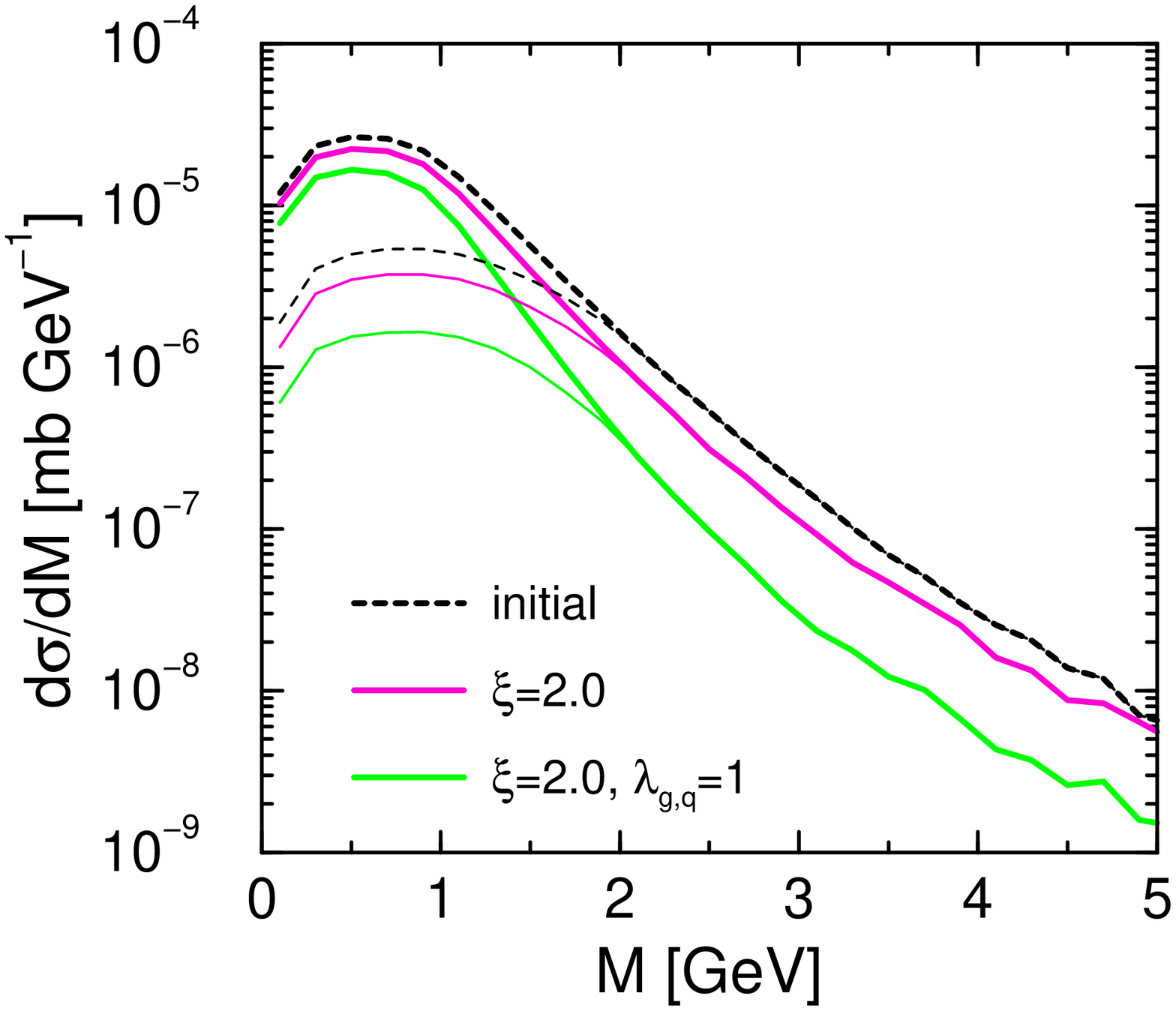,width=160pt,angle=-90}
\caption{Left panel: Single electron spectra
from semileptonic open charm decays
for the reaction Au + Au at $\sqrt{s_{NN}} = 200$ GeV for various
assumptions on the energy loss 
parameterized by the strenght factor $\xi$ as in Ref.\protect\cite{Gallmeister}
(dashed curve: no energy loss; magenta curve:
maximum temperature 550 MeV and initial quark-gluon undersaturation;  
green curve: the same maximum temperature but initial saturation). 
Data (non-photonic sources) from Ref.\protect\cite{PHENIX1}.
Right panel: Corresponding dilepton spectra within the PHENIX acceptance
with minimum single electron momentum of 500 MeV/c (solid curves) or 
1 GeV/c (thin curves).}
\label{fig6}
\end{figure*}

As in the first reference of Ref.\cite{Gallmeister} the gluon-radiative energy loss
of charmed quarks is parameterized in a geometrical model and the decay
single-electron spectra are compared
with data in Fig.~\ref{fig6} for heavy-ion collisions Au + Au
at $\sqrt{s_{NN}} = 200$ GeV.
Similar to the results at 130 GeV\cite{Gallmeister}, the
energy loss becomes noticeable and thus measurable 
at large momenta, where, however, precision data are still lacking. 
While the effect of the energy loss for the initial conditions with
undersaturated quark-gluon medium is hardly visible, 
the changes due to energy loss
in a saturated quark-gluon fluid are somewhat larger, but still not very strong.
We would like to add that these energy loss scenarios are probably
an upper limit, since we did not include the dead cone effect\cite{Ronny}
and neglected $log$ terms.
The di\-electron spectrum is more sensitive to the energy loss of charm quarks,
see Fig.~\ref{fig6}. The forthcoming dilepton data, therefore, may be essential
to pin down quantitatively the energy loss of charm quarks.      

\section{Summary}

At CERN-SPS the thermal electromagnetic radiation (dileptons and real photons)
off the fireball of strongly interacting matter
has been identified in heavy-ion collisions. 
It is compatible with the assumption of achieving maximum
temperatures of ${\cal O} (200)$ MeV, i.e.\ above the deconfinement temperature. 
The low-mass dilepton spectrum needs
a drastic reshaping in the resonance region, compatible with expectations
from chiral restoration. The first electromagnetic signals in heavy-ion
experiments at RHIC are compelling but need higher precision to arrive
at firm conclusions and to compare with the many existing predictions.
The important role of heavy flavor dynamics
seems to be confirmed. Besides dileptons and photons, diphotons should
be mentioned as interesting probe\cite{diphotons}.   
 
\section*{Acknowledgments}

This work is supported in part by BMBF 06DR121 and GSI.


\end{document}